\documentclass[a4paper,11pt]{article}
\usepackage{pos}

\usepackage{epstopdf}

\def\bea{\begin{eqnarray}}
	\def\eea{\end{eqnarray}}

\def\pp{\mbox{$p$-$p$}}

\def\auau{\mbox{Au-Au}}

\def\pbpb{\mbox{Pb-Pb}}

\def\aa{\mbox{A-A}}
\def\nn{\mbox{N-N}}

\def\pt{$p_t$}
\def\mt{$m_t$}
\def\yt{$y_t$}

\def\nch{$n_{ch}$}

\title{Azimuth Quadrupole Spectra derived from\\ 2.76 TeV Pb-Pb PID Differential $\bf v_2(p_t)$ Data}
 \ShortTitle{Azmuth Quadrupole Spectra from Differential $v_2(p_t)$ Data}

\author[a]{Thomas A. Trainor}

\affiliation[a]{University of Washington, Seattle, USA}

\emailAdd{ttrainor99@gmail.com}

\abstract{
	$v_2(p_t)$ data are intended to estimate the amplitude of an azimuth component of particle spectra interpreted as representing elliptic flow of a dense QCD medium.  As defined, $ v_2(p_t)$ is a ratio with a single-particle spectrum appearing in its denominator. Its numerator represents a spectrum Fourier component arising from a boosted particle source. The Cooper-Frye (CF) formalism may be used to describe emission from a boosted source. CF analysis reveals that the $v_2(p_t)$ numerator includes a factor $ p_t$ in the boost frame with major consequences for data interpretation. 
	A unique quadrupole $p_t$ spectrum may be isolated from $v_2(p_t)$ data and compared directly with the single-particle spectrum in the $\bf v_2$ denominator and with hydro theory. A monopole boost (aka radial flow) value may be estimated from $v_2(p_t)$ data. 
	Several novel results emerge via the Cooper-Frye analysis.
}

\FullConference{Proceedings of the Corfu Summer Institute 2025 "School and Workshops on Elementary Particle Physics and Gravity" (CORFU2025)\\
27 April - 28 September, 2025\\
Corfu, Greece\\}

\begin{document}
\maketitle


\section{Introduction} \label{intro}

Observation of evidence for {\em elliptic flow} within angular correlations on $(\eta,\phi)$ in the form of statistical measure $v_2(p_t)$ has been viewed as key evidence for formation of a quark-gluon plasma in heavy ion collisions at the relativistic heavy ion collider (RHIC) and large hadron collider (LHC)~\cite{olli,keystone}. Certain aspects of $v_2(p_t)$ data have been interpreted to indicate that a QGP has a low viscosity~\cite{rom,luzrat} leading to its interpretation as a ``perfect liquid''~\cite{perfect,qgp1,qgp2}. However, based on detailed studies of $v_2(p_t)$ in the context of precision measurements of 2D angular correlations~\cite{axialci,anomalous,ppquad} and a two-component model (TCM) of hadron production~\cite{ppprd,ppbpid,ppquad}, conventional methods associated with measure $v_2(p_t)$ and interpretation of its data trends may be questioned~\cite{v2ptb,njquad,quadspec}. 

This talk considers several related questions: What is the algebraic structure of $v_2(p_t)$ and what information is carried by corresponding data? What is the simplest parametrization of $v_2(p_t)$ data? What is the nature of the particle source and its motion that have lead to elliptic flow inferences. Is hydrodynamic (hydro) theory consistent with observations in those terms?

In response to those questions the structure of $v_2(p_t)$ is explored in the context of the Cooper-Frye formalism~\cite{cooperfrye}. A previous analysis of 
$v_2(p_t)$ data from 200 GeV \auau\ collisions that infers {\em quadrupole spectra}~\cite{quadspec} is reviewed. A similar analysis is applied to $v_2(p_t)$ data from 2.76 TeV \pbpb\ collisions. Quadrupole spectra from the two collision systems are quite similar. Trends for a {\em quadrupole amplitude} (in terms of number of correlated pairs) on event multiplicity \nch\ and collision energy $\sqrt{s_{NN}}$ for \pp\ as well as \aa\ collisions are considered, with a surprising result. Quadrupole data vs hydro theory is reexamined in that larger context.  This talk summarizes recent results reported in Ref.~\cite{tomquadnew}

.
\section{Elliptic flow, $\bf v_2(p_t)$ and the Cooper-Frye formalism} \label{cf}

As conventionally defined, elliptic flow measure $v_2(p_t)$ is in effect a ratio
\bea \label{v2pt}
v_2(p_t) &=& \frac{V_2\{\text{2D}\}(p_t) + \text{jet contribution (nonflow)}}{[(N_{part}/2) S_{NN}(p_t) + N_{bin}r_{AA}(p_t) H_{NN}(p_t)] \approx \bar \rho_0},
\eea
where the numerator is derived from 2D (two-dimensional) angular correlations on $(\eta,\phi)$ and the denominator is single-particle (SP) spectrum $ \bar \rho_0(p_t)$. $V_2\{\text{2D}\}(p_t)$ is derived from model fits to 2D angular correlations that effectively exclude jet contributions~\cite{njquad,noelliptic,multipoles} (and see Sec.~\ref{ppquad}). The second term is a possible bias arising from including some unwanted ``nonflow'' contributions depending on the $v_2$ method implemented. As defined, this conventional measure presents difficulties of interpretation.

As an exercise an ideal case may be considered -- the particle spectrum of a boosted source with known boost distribution $\Delta y_t(\phi_r)$ ($\phi_r = \phi - \Psi_r$, event plane angle) represented by
\bea  \label{boostspec}
\bar \rho_{2i}(y_{ti},\phi_r) & \propto & \exp\left\{-(m_i/T_2)[\cosh(y_{ti} - \Delta y_t(\phi_r)) - 1]\right\},
\eea
where transverse rapidity for hadron species $i$ is $y_{ti} = \ln[(m_{ti}+p_t)/m_i]$.
$\bar \rho_{2i}(y_{ti})$ is a Boltzmann exponential on transverse mass \mt\ in the boost frame boosted to the lab frame by $\Delta y_t(\phi_r)$ that includes monopole boost $\Delta y_{t0}$ ($\sim$radial flow) and quadrupole boost $\Delta y_{t2}$ ($\sim$elliptic flow). This is consistent with the Cooper-Frye approach~\cite{cooperfrye} albeit with simpler formulation.
The corresponding term in the numerator of Eq.~(\ref{v2pt}) is obtained, via Taylor expansion, as
\bea  \label{v2int}
V_{2i}(y_{ti}) &\equiv& \frac{1}{2\pi}\int_{-\pi}^\pi d\phi_r \bar \rho_{2i}(y_{ti},\phi_r)\cos(2\phi_r) \approx \bar \rho_{0i}(y_{ti})v_{2i}(y_{ti})
\\
\nonumber
&\approx& p_t'\frac{\Delta y_{t2}}{2T_2}\bar \rho_{2i}(y_{ti},\Delta y_{t0}),
\eea
where the first factor at right in the second line is \pt\ in the boost frame denoted by $p'_t$. Equation~(\ref{v2int}) may be rearranged to obtain
\bea  \label{quadspec}
\frac{\Delta y_{t2}}{2T_2}\bar \rho_{2i}(y_{ti},\Delta y_{t0}) \approx \frac{1}{p_t'} \bar \rho_{0i}(y_{ti})v_{2i}(y_{ti}),
\eea
where $\bar \rho_{2i}(y_{ti},\Delta y_{t0})$, the azimuth average of $\bar \rho_{2i}(y_{ti},\phi_r)$ in Eq.~(\ref{boostspec}), is a {\em quadrupole spectrum} defined in terms of measured quantities on the right side~\cite{quadspec}. In what follows, that formalism is applied to $v_2(p_t)$ data for two collision systems to explore boosted-source properties.

\section{200 GeV quadrupole spectra} \label{200}

As an introduction to quadrupole spectra this section summarizes  analysis, reported in Ref.~\cite{quadspec}, of identified-particle (PID) $v_2(p_t)$ data from 200 GeV \auau\ collisions.

\begin{figure}[h]
	\includegraphics[width=1.46in,height=1.32in]{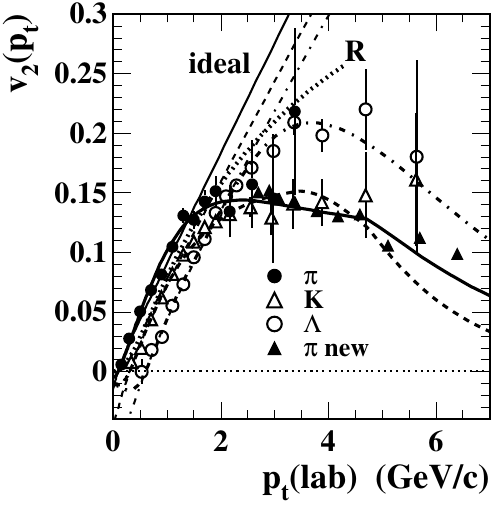}
	\includegraphics[width=1.46in,height=1.32in]{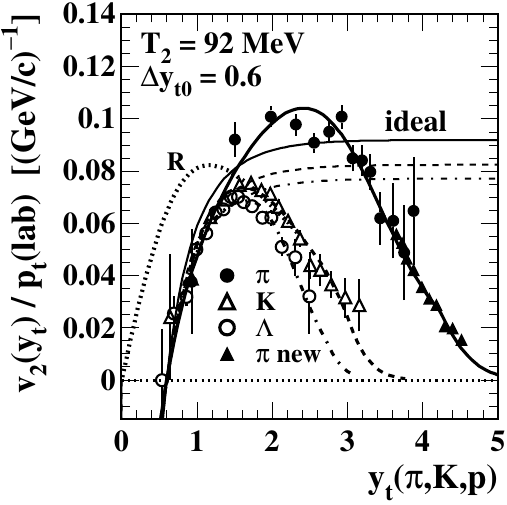}
	\includegraphics[width=1.46in,height=1.3in]{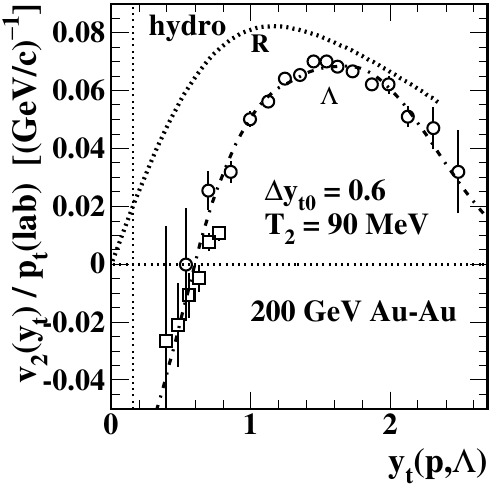}
	\includegraphics[width=1.46in,height=1.3in]{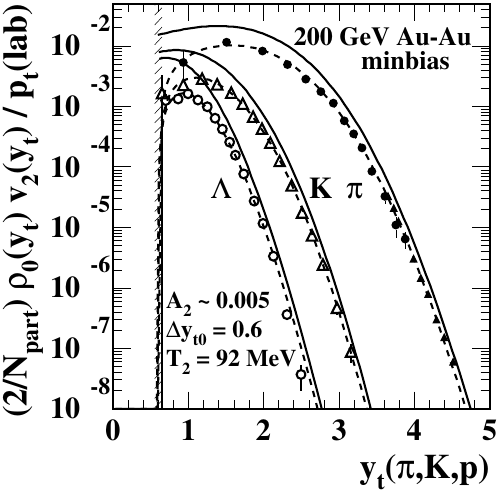}
	\caption{
		(a) $v_2(p_t)$ in conventional format.
		(b) Revised format with common intercept at $\Delta y_{t0}$.
		(c) Emphasis on Lambdas.
		(d) Quadrupole spectra in the lab frame.
	}
	\label{fig1}     
\end{figure}

Figure~\ref{fig1} (a) shows published data in a conventional plot format, $v_2(p_t)$ vs \pt, for pions, charged kaons and protons (points). The bold curves through data are described below. The thin curves extending off the panel are ``ideal hydro'' trends inferred from Eq.~(\ref{v2int}). Curve R is a viscous hydro prediction~\cite{rom}. New pion data (triangles) have been recently added~\cite{newstarpion}.

Figure~\ref{fig1} (b) shows the same data divided by \pt\ in the lab frame (as an initial approximation to $p'_t$ in the boost frame) and plotted vs proper \yt\ for each hadron species. Several important results emerge from this simple transformation. The data trends converge to a common zero intercept at $y_t \approx 0.6$ which may be interpreted as an estimate for monopole boost $\Delta y_{t0}$. The ``ideal hydro'' curves originate from the same intercept by construction but approach constant values at higher \yt. Viscous hydro curve R originates from $y_t = 0$ and is thus falsified by data.

Figure~\ref{fig1} (c) shows an expanded view of (b) for Lambda data with updated higher-statistics data at lower \yt\ (open boxes)~\cite{newlambda}. This plot emphasizes the difference between $v_2(p_t)$ data and hydro theory prediction R that appears to have a monopole boost {\em distribution} (Hubble-like medium expansion?) different from that required by data (Lambda zero intercept at $y_t \approx 0.6$).

Figure~\ref{fig1} (d) shows data in the form of Eq.~(\ref{quadspec}) but with \pt\ in the lab as an approximation (points). The required SP spectra $\bar \rho_{0i}(y_t)$ are as in Fig.~\ref{fig6} (a,b)~\cite{hardspec}. The dashed (with \pt\ in lab frame) and solid (with \pt\ in boost frame) curves are derived from the quadrupole spectrum (solid curve) in Fig.~\ref{fig7} (b) back transformed to plotting format (d) and also to solid, dashed and dash-dotted curves in panels (a,b,c). The parameters $(T_2,n_2)$ that define the solid curve in Fig.~\ref{fig7} (b) then also describe $v_2(p_t)$ data in Fig.~\ref{fig1} (a,b,c,d) formats via $\Delta y_{t0}$ and $\bar \rho_{0i}(y_t)$.



\section{2.76 TeV Pb-Pb $\bf v_2(p_t)$ data} \label{276}

Following sections describe similar quadrupole spectrum analysis of $v_2(p_t,n_{ch})$ data from 2.76 TeV \pbpb\ collisions for six event classes~\cite{alicev2ptb}.

\begin{figure}[h]
	\includegraphics[width=2.92in,height=1.4in]{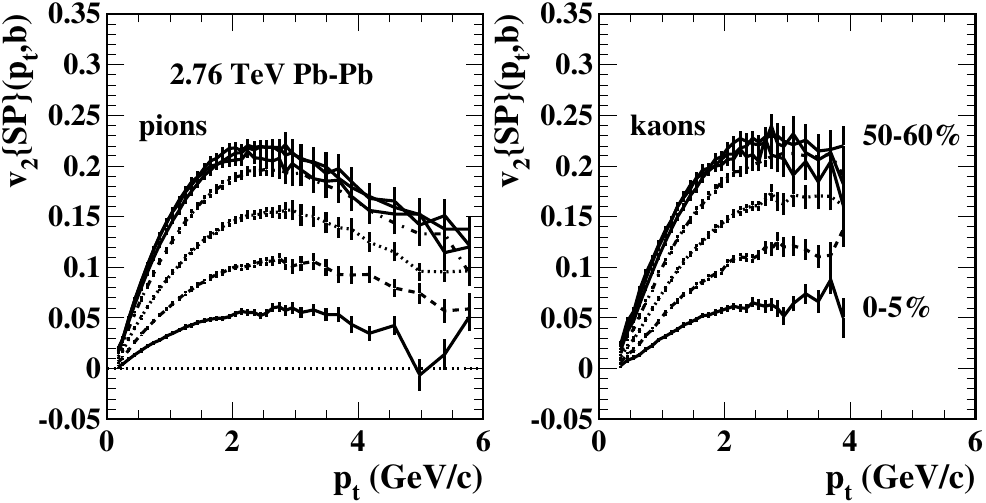}
	\includegraphics[width=2.92in,height=1.4in]{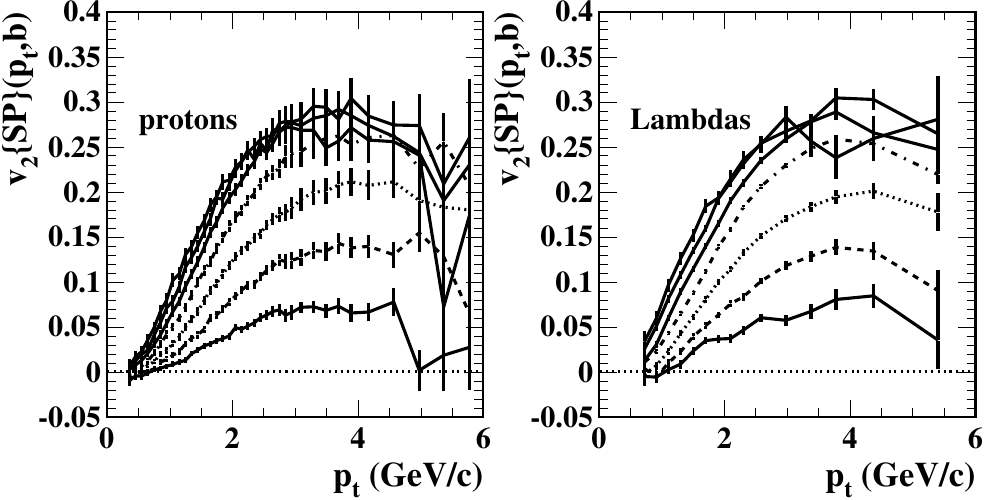}
	\caption{$v_2(p_t,n_{ch})$ from 2.76 TeV \pbpb\ collisions for (a) pions, (b) charged kaons, (c) protons and (d) Lambdas as published.
	}
	\label{fig2}     
\end{figure}

Figure~\ref{fig2} shows $v_2(p_t,n_{ch})$ data (points) in a conventional format vs linear \pt\ for (a) pions, (b) charged kaons, (c) protons and (d) Lambdas. In this format data uncertainties (statistical and systematic combined in quadrature) vary greatly with increasing \pt\ from negligible at lowest \pt\ to a fair fraction of the plotting space at highest \pt. The trend hints that important information at low \pt\ (where most jet fragments reside) may be strongly suppressed visually.

The spacing of event classes on charge multiplicity \nch\ is also problematic. The most-peripheral event class above is assigned as 50-60\% which leaves a large interval of more-peripheral collisions unexamined. As suggested by Fig.~\ref{fig3} (a), the ability of some conventional $v_2$ analyses methods (e.g.\ $v_2\{4\}$) to cope with lower A-B event multiplicities appears quite limited, especially as compared to model fits to 2D angular correlations (see Sec.~\ref{ppquad}). The cost of such limitations becomes apparent in Fig.~\ref{fig9} (b) as noted below.

\section{Particle source boost variation with event $\bf n_{ch}$} \label{boost}

The analysis described here is directed to obtain as differential an examination of $v_2(p_t,n_{ch})$ data as is possible. The data are rescaled by \pt-integral $v_2(b) \leftrightarrow v_2(n_{ch})$ to examine shape evolution with event class, then replotted in the format of Fig.~\ref{fig1} (b) to examine boost variation with \nch\ (not accessible with minimum-bias 200 GeV data in Sec.~\ref{200}).

\begin{figure}[h]
	\includegraphics[width=2.92in,height=1.4in]{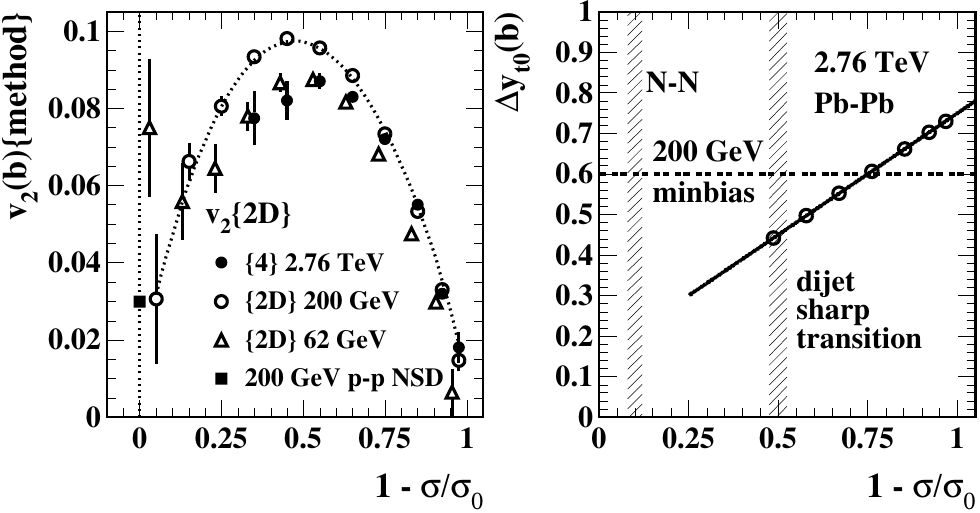}
	\includegraphics[width=2.92in,height=1.4in]{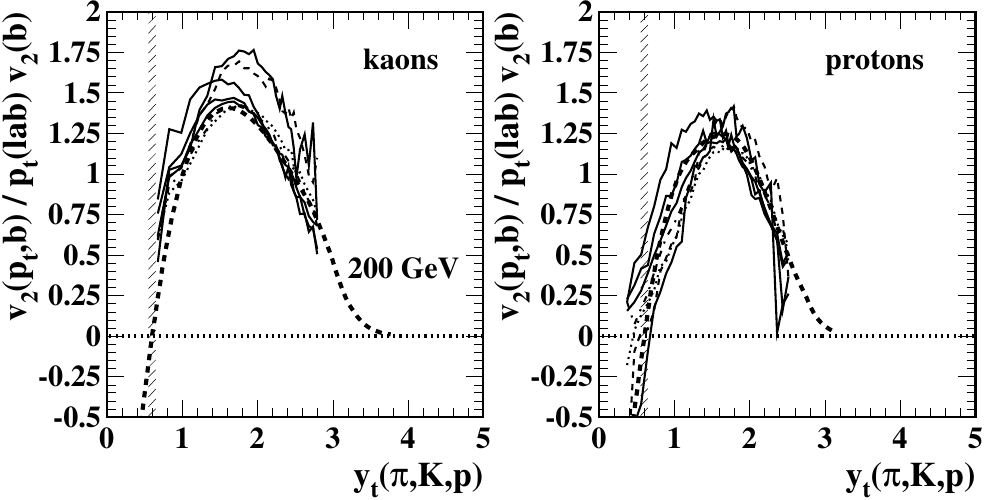}
	\caption{
		(a) $v_2(b)$ for 2.76 TeV \pbpb\ collisions (solid dots) and model fits to 2D angular correlations for 62 and 200 GeV \auau\ collisions (open symbols) and 200 GeV \pp\ collisions (solid square).
		(b) Inferred centrality variation of monopole boost $\Delta y_{t0}$ (points) for 2.76 TeV \pbpb\ collisions.
		Rescaled $v_2(b)$ data for 2.76 TeV \pbpb\ collisions showing variation of $\Delta y_{t0}$ with centrality for kaons (c) and protons (d).
	}
	\label{fig3}     
\end{figure}

Figure~\ref{fig3} (a) shows  \pt-integral $v_2(b)$ data for 2.76 TeV \pbpb\ (solid dots) and for 200 and 62 GeV \auau\ data (open symbols), where notation \{X\} denotes analysis method X: model fits to 2D angular correlations (2D) or four-particle cumulants (4). Also included is a value for non-single-diffractive (NSD) 200 GeV \pp\ collisions (solid square) as described in Sec.~\ref{ppquad}.

Figure~\ref{fig3} (c,d) shows $v_2(p_t,n_{ch})$ data in the format of Fig.~\ref{fig1} (b) that reveal significant variation of monopole boost $\Delta y_{t0}$ (zero intercepts on \yt) with values estimated in panel (b). In what follows, $v_2(p_t)$ data are shifted on \yt\ to coincide with a common boost $\Delta y_{t0} = 0.6$.

\begin{figure}[h]
	\includegraphics[width=2.92in,height=1.4in]{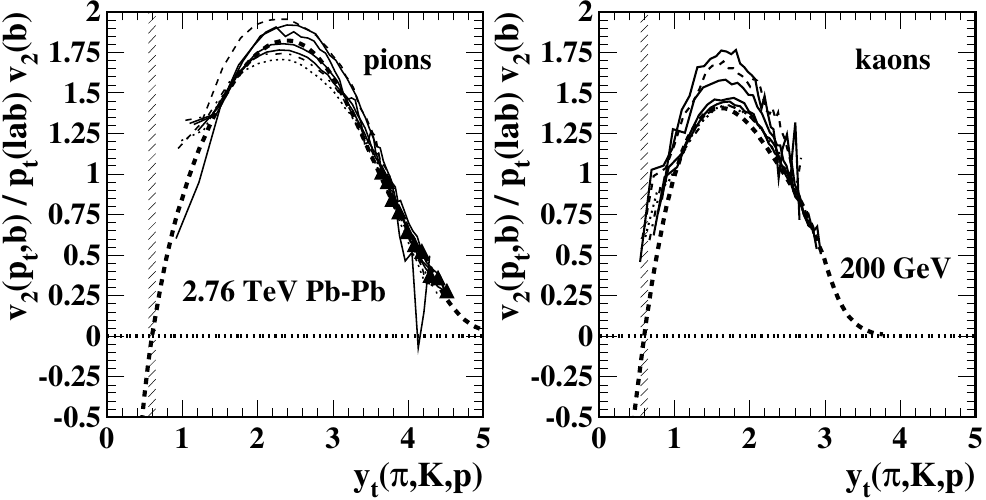}
	\includegraphics[width=2.92in,height=1.4in]{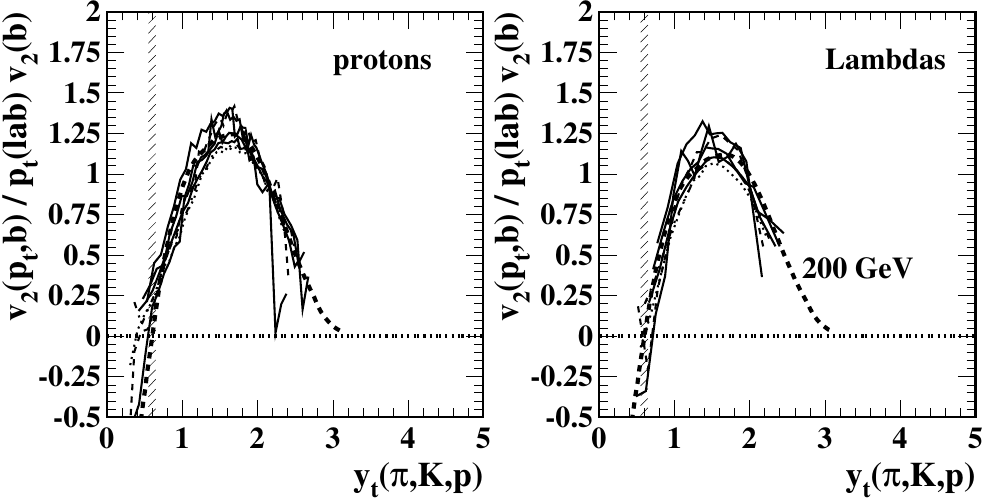}
	\caption{Rescaled 2.76 TeV \pbpb\ $v_2(p_t)$ data in the format of Fig.~\ref{fig1} (b) shifted to common $\Delta y_{t0}$ for (a) pions, (b) charged kaons, (c) protons and (d) Lambdas.
	}
	\label{fig4}     
\end{figure}

Figure~\ref{fig4} shows $v_2(p_t,n_{ch})$ data for pions, charged kaons, protons and Lambdas shifted on \yt\ as described above and rescaled by \pt-integral $v_2(n_{ch})$ per Fig.~\ref{fig3} (a) (solid dots). Several event classes coincide approximately with the results for 200 GeV (dashed curves, see Fig.~\ref{fig1}).


\section{Quadrupole spectra on transverse rapidity $\bf y_{ti}$} \label{quadyt}

In order to derive quadrupole spectra from $v_{2i}(p_t,n_{ch})$ data, single-particle spectra $\bar \rho_{0i}(p_t,n_{ch})$ are required per Eq.~(\ref{quadspec}) as defined on \pt\ values for $v_2(p_t)$ data.

\begin{figure}[h]
	\includegraphics[width=2.92in,height=1.4in]{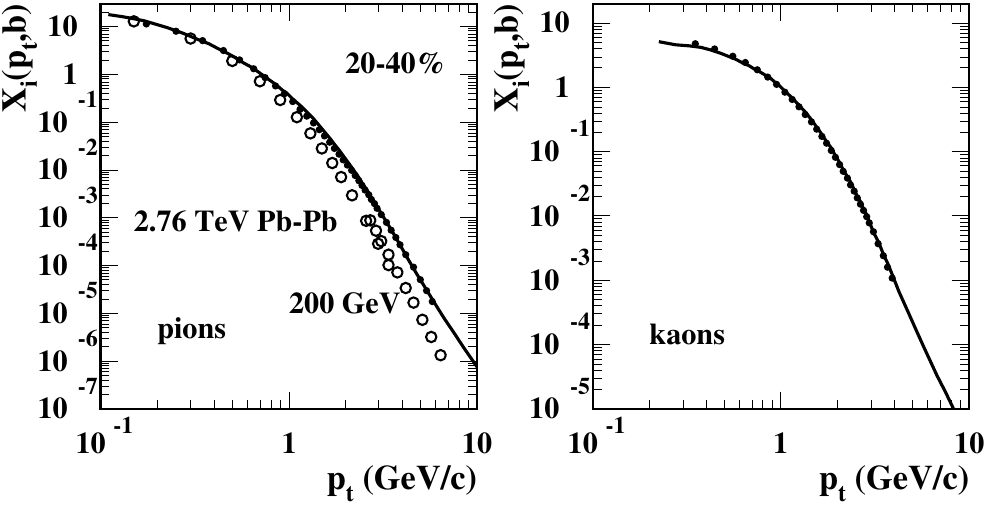}
	\includegraphics[width=2.92in,height=1.4in]{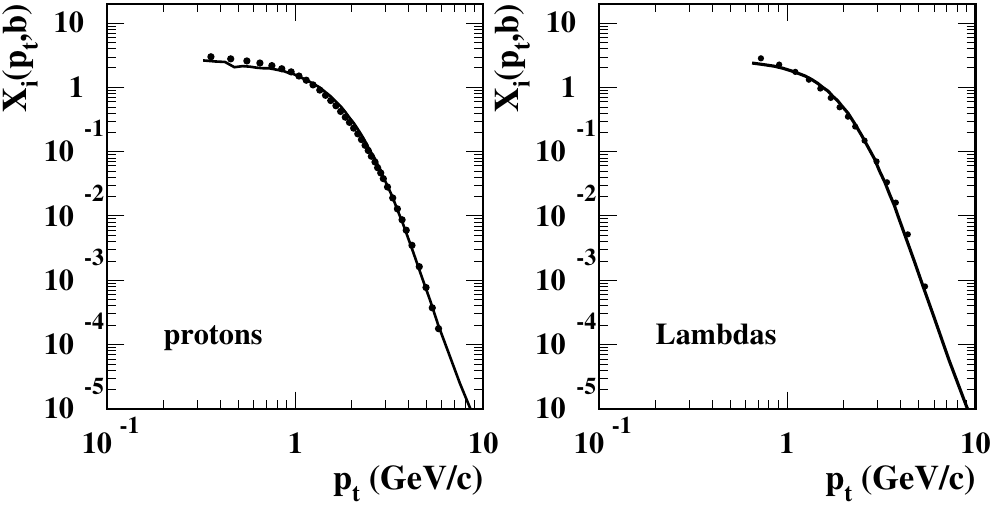}
	\caption{Soft-rescaled SP spectra $X_i(p_t)$ (solid curves) and models defined on $v_2(p_t)$ \pt\ values (solid points) for 
		(a) pions,
		(b) charged kaons,
		(c) protons and
		(d) Lambdas The open symbols in (a) are a \auau\ model for Fig.~\ref{fig1} (d) for comparison.
	}
	\label{fig5}     
\end{figure}

\begin{figure}[h]
	\includegraphics[width=1.46in,height=1.439in]{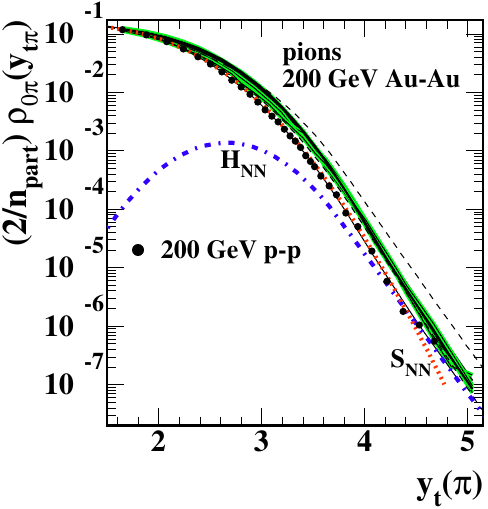}
	\includegraphics[width=1.46in,height=1.439in]{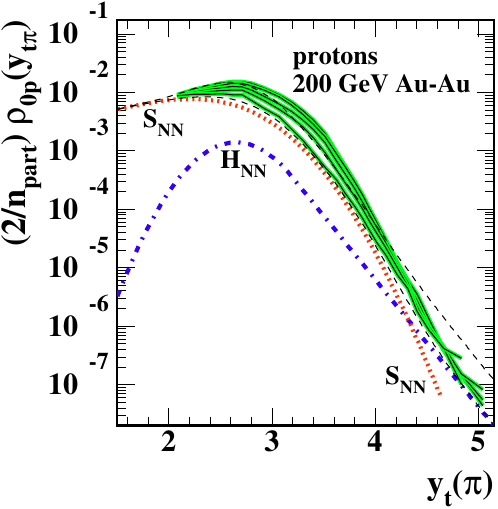}
	\includegraphics[width=2.92in,height=1.4in]{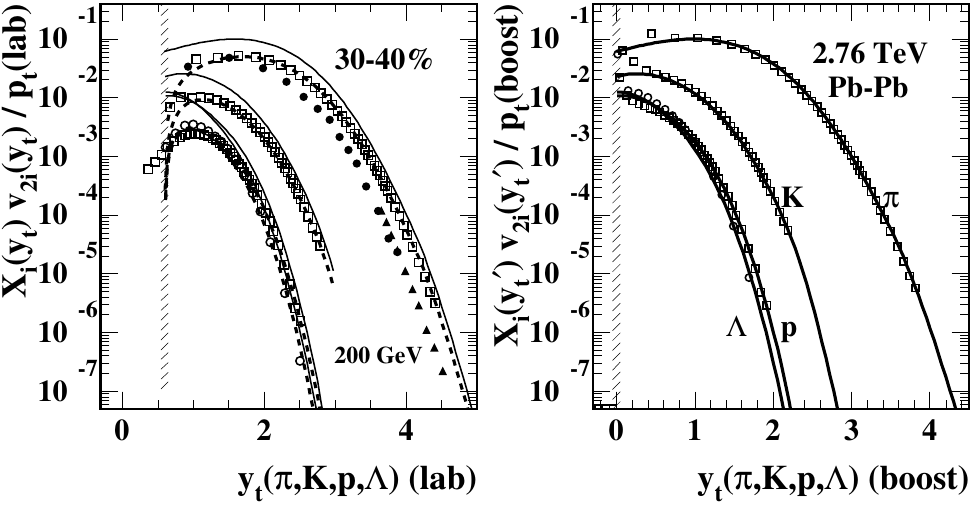}
	\caption{SP spectra (densities on \yt) for (a) pions and (b) protons from 200 GeV \auau\ collisions. Quadrupole spectra for pions, kaons, protons and Lambdas (points) (c) in the lab frame and (d) in the boost frame. (c,d) curves are explained in the text.
	}
	\label{fig6}     
\end{figure}

Figure~\ref{fig5} shows hadron spectra (solid curves) as reported in Ref.~\cite{alicepbpbpidspec} for 20-40\% central \pbpb\ collisions. Since there is little variation in rescaled $v_2(p_t)$ data as it appears in Fig.~\ref{fig4} the 20-40\% central data are chosen as being least susceptible to jet bias (``nonflow''). The solid dots for four hadron species are model spectra defined on $v_2$ data \pt\ values that approximate the data SP spectra reasonably well. Also included is a spectrum model for 200 GeV pion analysis (open circles) as required in Fig.~\ref{fig1} (d) for comparison. Comparable spectrum data for 200 GeV \auau\ (as densities on \yt) are shown in Fig.~\ref{fig6} (a,b)~\cite{hardspec}.


Figure~\ref{fig6} (c) shows quadrupole spectra as $X_i(y_t) v_{2i}(y_t) / p_t(\text{lab})$ (open points), where a SP spectrum rescaled by corresponding soft density $\bar \rho_{si}$ is denoted by  $X_i(y_t) \equiv \bar \rho_{0i} / \bar \rho_{si}$. The dashed curves represent Eq.~(\ref{quadspec}) but with $p_t'$ (in boost frame) replaced by \pt\ (in lab frame). The solid curves are copied from panel (d).


Figure~\ref{fig6} (d) shows quadrupole spectra as $X_i(y_t) v_{2i}(y_t) / p_t(\text{boost})$ per Eq.~(\ref{quadspec}) as written. Solid points in panel (c) are 200 GeV pion results from Fig.~\ref{fig1} (d) rescaled to match 2.76 TeV pion data at low \yt. The final step is transformation to transverse mass \mt\ in the boost frame.

\section{Quadrupole spectra on transverse mass $\bf m_{ti}$ in the boost frame} \label{quadmt}

Described above is a sequence of {\em homeomorphisms}. Its final step is transformation to transverse mass in the boost frame of particle source(s). Intermediate spectra on $y_{ti}$ in the boost frame in Fig.~\ref{fig6} (d) are transformed to  transverse mass $m_{ti}$ via the Jacobian $y_{ti}/(m_{ti}-m_t)p_t$.

\begin{figure}[h]
	\includegraphics[width=2.92in,height=2.645in]{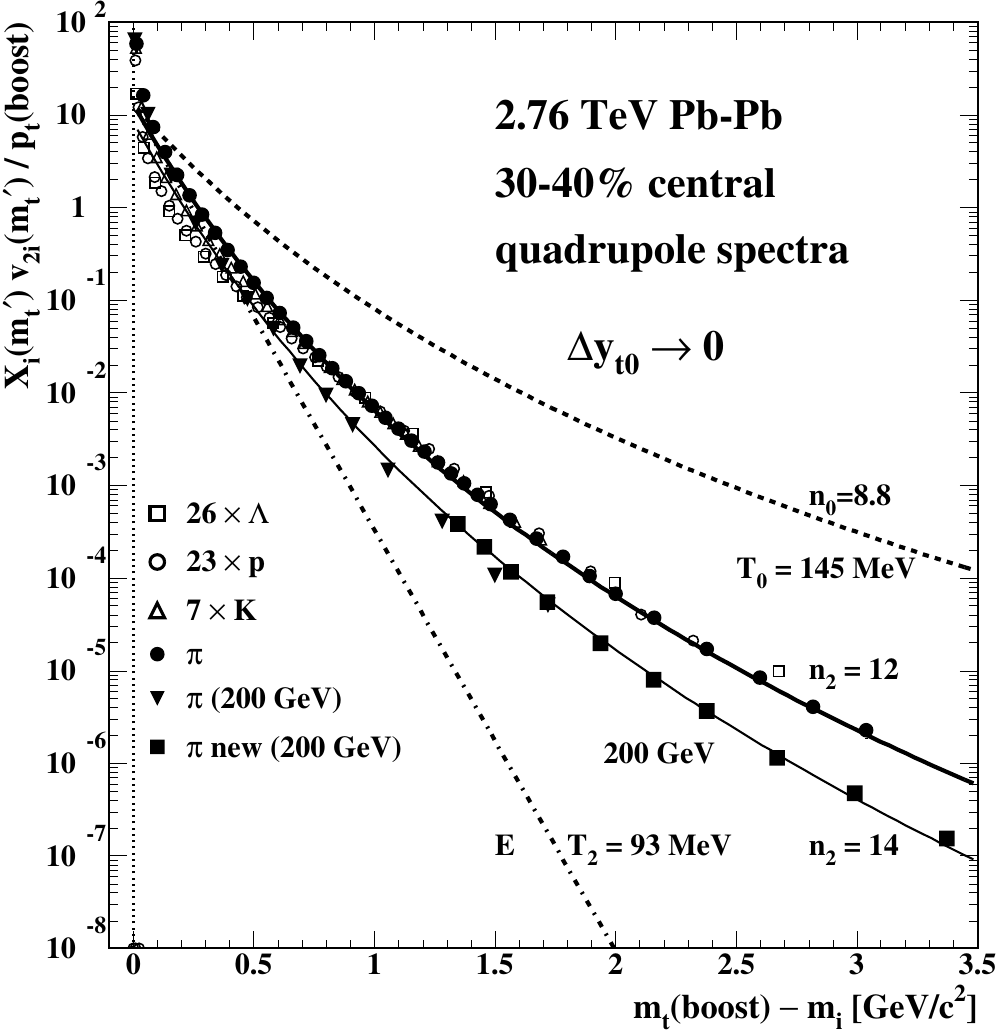}
	\includegraphics[width=2.92in,height=2.6in]{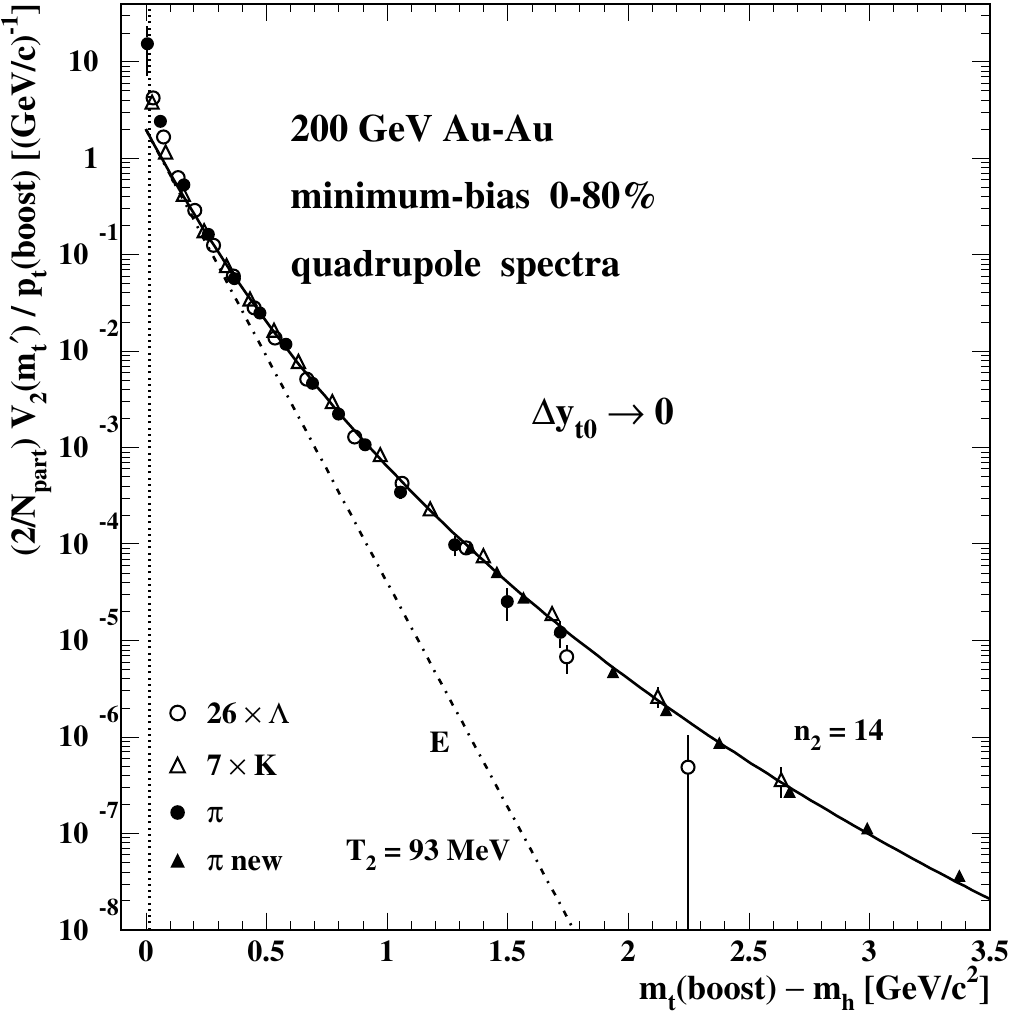}
	\caption{Quadrupole spectra as densities on transverse mass \mt\ in the boost frame (points) for (a) 2.76 TeV \pbpb\ collisions and (b) 200 GeV \auau\ collisions. Panel (a) includes 200 GeV  results from (b) for comparison.
	}
	\label{fig7}     
\end{figure}

Figure~\ref{fig7} (a) shows quadrupole spectra for four hadron species (bold solid curve and associated points) from 30-40\% central 2.76 TeV \pbpb\ collisions. The bold solid curve is defined by two parameters: slope parameter $T_2 = 93$ MeV and power-law tail exponent $n_2 = 12$. That single curve is back-transformed to the solid or dashed curves in Fig.~\ref{fig6} (c,d) describing data. The dashed curve is soft-component model $\hat S_0(m_t)$ for SP spectra that demonstrates the large difference between quadrupole spectra and SP spectra.

Figure~\ref{fig7} (b) shows quadrupole spectra for 200 GeV \auau\ collisions corresponding to the presentation in Sec.~\ref{200}. Those 200 GeV results for pions, rescaled to match 2.76 TeV results at low \pt, are copied in panel (a) as the thin solid curve and associated data. The only significant difference is a larger value for power-law exponent $n$ (softer spectrum) for the lower energy.

\section{Quadrupole amplitude $\bf V_2^2$ from p-p angular correlations} \label{ppquad}

Analysis of 2D angular correlations via model fits was pioneered at  RHIC by  early studies~\cite{axialci,axialcd,v2ptb}. That technique enables precise isolation of several correlated-pair sources, including longitudinal projectile-nucleon dissociation (soft), Bose-Einstein correlations, parton fragmentation to jets (hard), and quadrupole-related production ($\cos(2\phi)$ sinusoid on $\phi$). Obtaining $v_2(p_t)$ data via 2D model fits excludes any significant bias from jet fragments (``nonflow'') and extends data to low event \nch\ for several hadron species. Reference~\cite{ppquad} reports quadrupole amplitudes (correlated-pair numbers) for seven \nch\ event classes from 200 GeV \pp\ collisions.

\begin{figure}[h]
	\includegraphics[width=1.46in,height=1.4in]{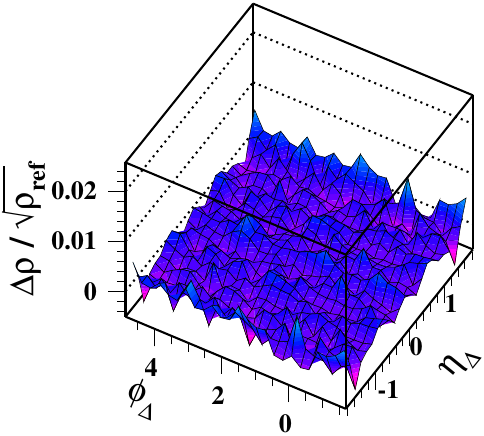}
	\includegraphics[width=1.46in,height=1.4in]{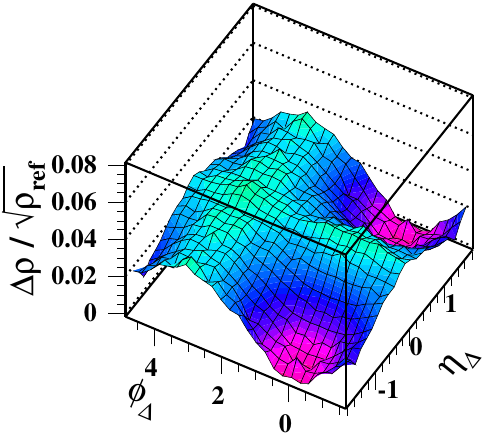}
	\includegraphics[width=1.46in,height=1.4in]{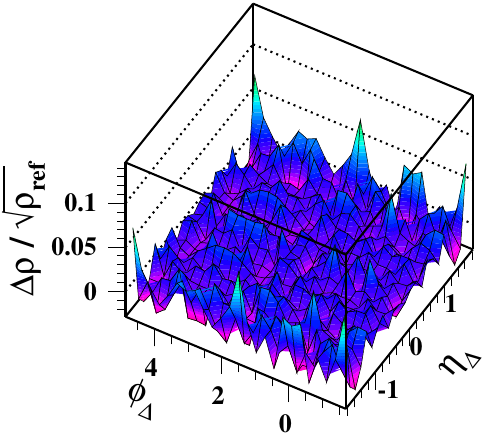}
	\includegraphics[width=1.46in,height=1.4in]{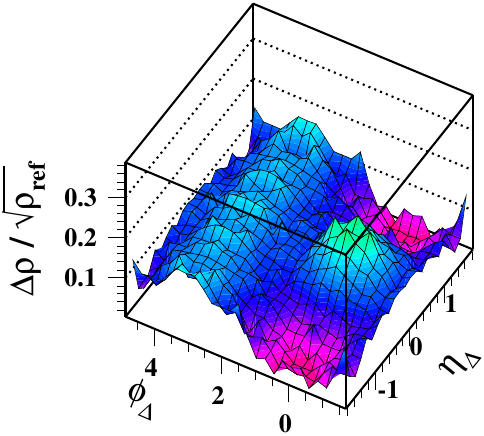}
	\caption{Results of model fits to 2D angular correlations from 200 GeV \pp\ collisions for low \nch\ (a,b) and high \nch\ (c,d) collision events, with (a,c) fit residuals and (b,d) jet and quadrupole elements. Quadrupole and jet components are comparable in (d).
	}
	\label{fig8}     
\end{figure}

Figure~\ref{fig8} shows 2D angular correlations for event class 1 (lowest \nch) (a,b) and event class 6 (of 7) (c,d). Panels (a,c) show fit residuals consistent with statistical uncertainties, i.e.\ all information is represented by the model. Panels (b,d) show a combination of hard-component pairs (same-side 2D Gaussian and away-side dipole) and cylindrical quadrupole $\cos(2\phi)$.

In panel (b) jet structure dominates correlations. The away-side dipole peaked at $\phi_\Delta = \pi$ is broad while the same-side 2D jet peak is strongly elongated on $\phi_{\Delta}$. In panel (d) the quadrupole element is quite apparent via the following features: (i) the away-side 1D peak radius of curvature is greatly reduced from panel (b) due to superposition of one lobe of the quadrupole element. (ii) Curvature at large $\eta_\Delta$ and $\phi_\Delta \approx 0$ is consistent with zero as opposed to the case for panel (b). (iii) The same-side jet peak appears strongly narrowed on $\phi_\Delta$, but the jet contribution has the same shape as in panel (b). The other quadrupole lobe, superposed on the jet peak, appears as the prominent structure. When properly understood, 2D correlation structure makes clear the large  contribution from the nonjet quadrupole element at high \nch.

\section{$\bf V_2^2$ charge-multiplicity and collision-energy dependence} \label{v22nch}

\begin{figure}[t]
	\includegraphics[width=1.46in,height=1.4in]{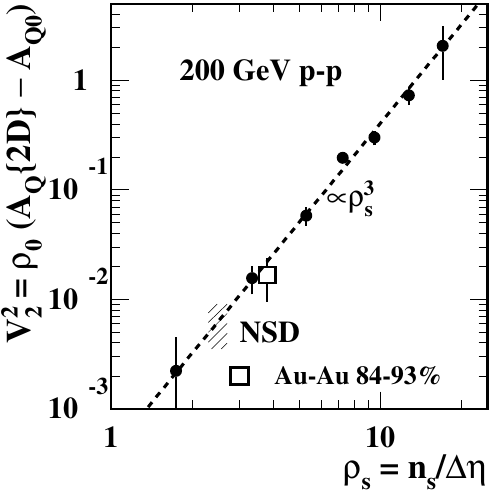}
	\includegraphics[width=1.46in,height=1.45in]{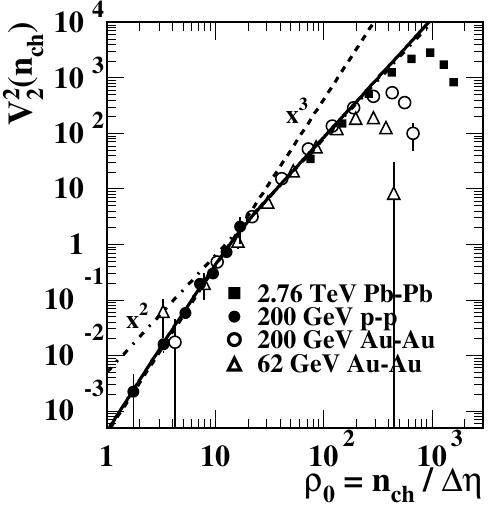}
	\includegraphics[width=1.46in,height=1.4in]{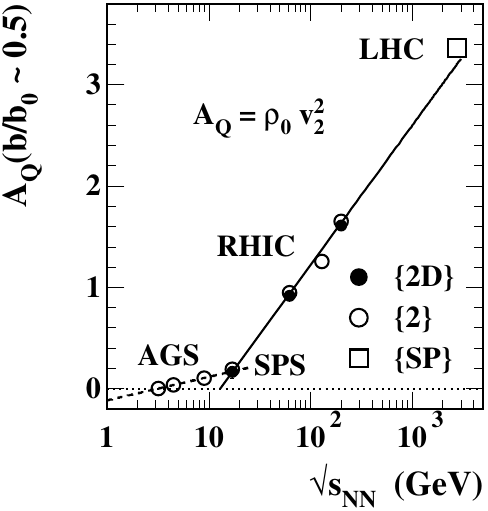}
	\includegraphics[width=1.46in,height=1.4in]{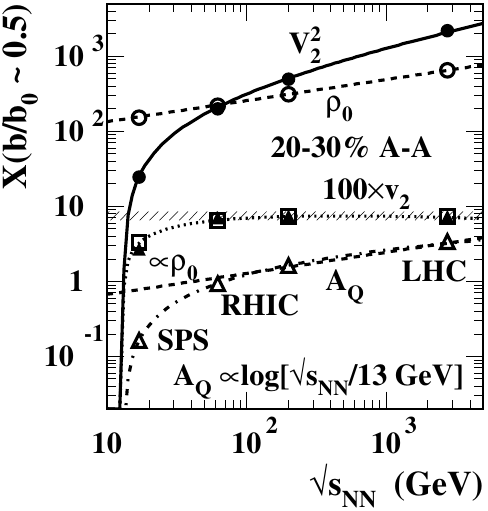}
	\caption{
		(a) Quadrupole amplitude $V_2^2$ vs charge-density soft component $\bar \rho_s$ for 200 GeV \pp\ collisions.
		(b) Quadrupole amplitude $V_2^2$ vs charge density $\bar \rho_0$ for 200 GeV \pp, 62 and 200 GeV \auau\ and 2.76 TeV \pbpb\ collisions.
		(c) Per-particle quadrupole amplitude $A_Q$ for \aa\ collisions vs collision  energy $\sqrt{s_{NN}}$ from AGS to LHC energies.
		(d) Energy dependence of several quadrupole-related measures.
	}
	\label{fig9}     
\end{figure}

Figure~\ref{fig9} (a) shows quadrupole amplitude $V_2^2(n_{ch}) \equiv \bar \rho_0^2 v_2^2(n_{ch})$ for 200 GeV \pp\ collisions plotted vs \nch\ soft component $n_s$ as reported in Ref.~\cite{ppquad}. The trend is a cubic relation $\propto n_s^3$ where $n_s$ may be interpreted to represent participant low-$x$ gluons at these higher collision energies. Jet and quadrupole features have comparable amplitudes in Fig.~\ref{fig8} (d) because of the quadratic (jet) vs cubic (quadrupole) trends on $n_s$. Given the observed relation $n_h \propto n_s^2$ for \pp\ collisions  the \pp\ correlation result may be summarized as $V_2^2(n_{ch}) \propto \bar \rho_s \bar \rho_h$.

Figure~\ref{fig9} (b) shows 62 and 200 GeV \auau\ $v_2\{\text{2D}\}(b)$ data from Fig.~\ref{fig3} (a) converted to $V_2^2(n_{ch})$ (open symbols) and plotted vs charge density $\bar \rho_0$. The \pp\ result from panel (a) (solid dots) is superposed with its cubic trend (dashed line). Based on the \pp\ data one may speculate that the same trend could persist for \aa\ collisions as 
\bea \label{v22aa}
V_2^2(n_{ch}) &\propto&  \bar \rho_s \bar \rho_h ~\approx~ (N_{part}/2) \bar \rho_{sNN} \times N_{bin} \bar \rho_{hNN},
\eea
a product of soft times hard components from the TCM representation of \aa\ particle densities. The solid curve in panel (b) is  Eq.~(\ref{v22aa}) based on TCM results reported in Ref.~\cite{tompbpb}. The solid curve is rescaled to best describe data at lower \nch. In Ref.~\cite{tompbpb} it is noted that for \aa\ collisions a range of peripheral collisions up to a transition point ($\bar \rho_0 \approx 15$) gives results consistent with single \nn\ collisions due to exclusivity~\cite{tomexclude}, wherein $N_{part}/2 \approx N_{bin} \approx 1$. Above the transition point densities $\bar \rho_{xNN}$ are approximately fixed, and geometry parameters $N_{part}/2$ and $N_{bin}$ vary in such a way that their product is approximately $\propto \bar \rho_0^2$. The solid curve is consistent with those trends (dashed and dash-dotted lines) {\em and} $V_2^2(n_{ch})$ data up to semicentral \aa\ collisions ($\bar \rho_0 \approx 200$). The same procedure applied to \pbpb\ data (solid squares)~\cite{alicepbpbyields} follows that trend. Falloffs from Eq.~(\ref{v22aa}) at highest \nch\ may be due to overlap of {\em multiple} quadrupole emissions within a single event averaged over $\phi_{\Delta}$ to reduced amplitude. Quadrupole {\em data} in the form $V_2^2(n_{ch})$ were {\em not rescaled} for this figure.

Figure~\ref{fig9} (c,d) shows the collision-energy dependence of the nonjet quadrupole. Panel (c) shows per-particle correlation amplitude $A_Q(\sqrt{s_{NN}})$ with $V_2^2 = \bar \rho_0 A_Q$~\cite{v2ptb}. Below a transition point near $\sqrt{s_{NN}} = 13$ GeV the amplitude transitions from negative to positive values, presumably corresponding to plastic flow of intact nucleons. Above the transition  data follow the trend $A_Q(\sqrt{s_{NN}}) \propto \log(\sqrt{s_{NN}} / 13 ~\text{GeV}$) from SPS to LHC energies~\cite{davidhq} which may correspond to a quadrupole production process based on three-gluon interactions as in Fig.~\ref{fig9} (a).

Figure~\ref{fig9} (d) shows the energy dependence of several quadrupole-related quantities for comparison. Open triangles are 
$A_Q(\sqrt{s_{NN}})$ repeated from panel (c). 
Open circles are $\bar \rho_0(\sqrt{s_{NN}})$.
Solid dots are $V_2^2(\sqrt{s_{NN}}) = \bar \rho_0(\sqrt{s_{NN}}) A_Q(\sqrt{s_{NN}})$.  
The dash-dotted curve is $\propto \log(\sqrt{s_{NN}} / 13 ~\text{GeV}$) from panel (c). The dashed line is empirical power law $\bar \rho_0 \propto \sqrt{s_{NN}}^{0.28}$. The dotted curve is an empirical $v_2(\sqrt{s_{NN}})$ trend inferred from the $A_Q$ and $\bar \rho_0$ trends, with open boxes as predictions for measured $v_2$. $v_2\{\text{2D}\}$ data, shown as solid triangles, agree well with the predictions.


\section{Mesons and baryons -- NCQ scaling} \label{ncq}

Reference~~\cite{alicev2ptb} reports on ``test[s] of scaling properties'' relating to its Figs.~8 and 10. The goal of ``scaling'' is apparently to reach a $v_2$ data configuration where ``...the various [PID] hadron species approximately follow a common behavior.'' It should become apparent from the present analysis and related studies that attempts at such scaling result from a misunderstanding of the nature of the $v_2$ measure and its relation to collision dynamics.

\begin{figure}[h]
	\includegraphics[width=2.92in,height=1.4in]{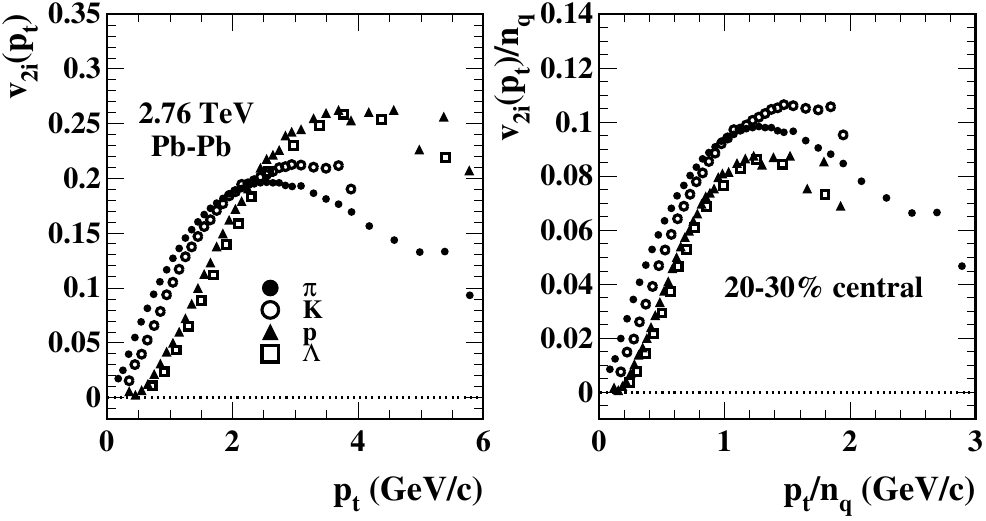}
	\includegraphics[width=2.92in,height=1.4in]{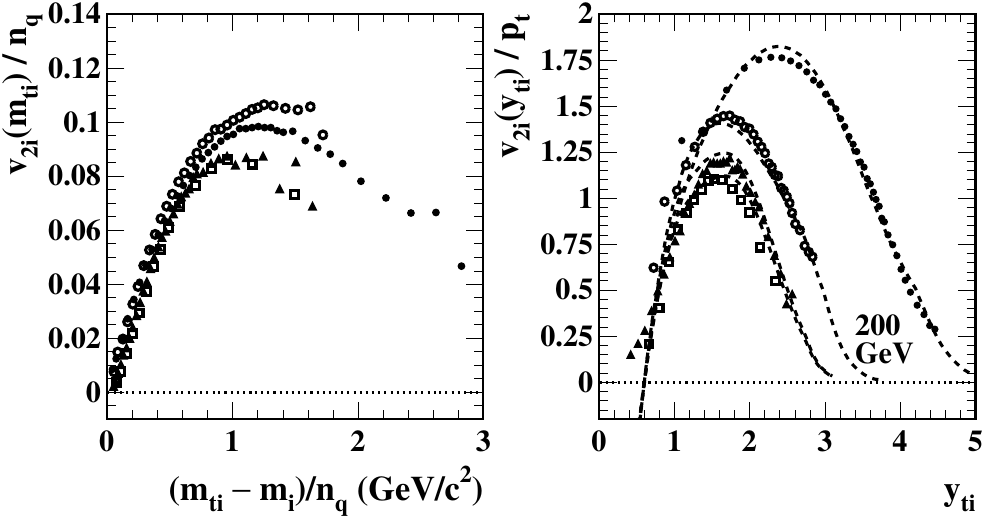}
	\caption{
		(a) Conventional format $v_2(p_t)$ vs \pt\ for pions, kaons, protons and Lambdas from 2.76 TeV \pbpb\ collisions.
		(b) Data in (a) replotted per ``constituent-quark'' $n_q$ scaling.
		(c) Data in (b) replotted per transverse ``kinetic energy'' $m_{ti} - m_i$.
		(d) Data in (a) plotted as $v_2(y_{ti}) / p_t(\text{lab})$ vs $y_{ti}$ showing a common intercept at  $\Delta y_{t0}$.
	}
	\label{fig10}     
\end{figure}

Figure~\ref{fig10} (a) shows \pbpb\ $v_2(p_t)$ data in the format of Fig.~\ref{fig1} (a) above. The data symbols are here made small enough to actually distinguish fine structure of the data trends, in contrast to Fig.~8 of Ref.~\cite{alicev2ptb}. The  vertical displacement between mesons and baryons at higher \pt\ has been a sustained point of interest with responding theoretical conjecture~\cite{rudy,duke,tamu}. Panel (b) shows one example of ``scaling'' in which horizontal and vertical variables are rescaled (divided) by ``a number of constituent quarks'' $n_q$ (NCQ scaling). The result is reduced horizontal separation among mesons and baryons on rescaled \pt\ at lower \pt\ and vertical separation at higher \pt\ which is the desired effect, but the meaning of such actions and results is not clear.

Figure~\ref{fig10} (c) shows an alternative scaling in that \pt\ is replaced by transverse ``kinetic energy'' $m_{ti} - m_t$. It is evident that such substitution brings the trends for different hadron species closer near the origin. However, the reason for such convergence is trivial. The displacements on \pt\ at lower \pt\ in panel (a) correspond to fixed monopole boost $\Delta y_{t0} \approx 0.6$, in which case since $p_t = m_i \sinh(y_{ti}) \approx m_i y_{ti}$ at low \pt\ zero intercepts on \pt\ are approximately at $m_i \Delta y_{t0}$. For panel (c), $m_{ti} - m_i = m_i[\cosh(y_{ti})-1] \approx m_i y_{ti}^2 / 2$. For $\Delta y_{t0} = 0.6$, $y_{ti}^2 / 2 \rightarrow 0.18$. Thus, separations among intercepts in (b) are reduced by factor 3 in (c). The point missed is that the essential element is monopole boost value $\Delta y_{t0} \approx 0.6$ which is clearly demonstrated in panel (d) given a plot format that responds to the algebraic structure of $v_2(p_t)$ described in Sec.~\ref{cf}. Once a fixed monopole boost is obtained the procedures presented in Sec.~\ref{200} and Ref.~\cite{quadspec} lead to Fig.~\ref{fig7}, where {\em transformed} spectra for several hadron species {\em do share a ``common behavior.''}

\section{$\bf v_2(p_t)$ vs hydrodynamic theory} \label{hydro}

Regarding comparison of $v_2(p_t)$ data to hydro theory, Ref.~\cite{alicev2ptb} asserts that ``It has been established that hydrodynamic as well as hybrid [hydro plus cascade] models...describe  the soft particle production at both RHIC and the LHC fairly well.'' The meaning of ``soft particle production'' is not specified but may refer to anything appearing below 2 GeV/c which is the conventional upper limit for ``soft'' particles. The assertion contradicts broad evidence that {\em most jet fragments} appear within that \pt\ interval~\cite{jetspec2}. There is also the issue of persistent major disagreement between viscous hydro predictions and $v_2(p_t)$ data as in Fig.~\ref{fig1} (c)

Comparisons in its Fig.~6 are made between Ref.~\cite{alicev2ptb} \pbpb\ data and the VISHNU Monte Carlo~\cite{vishnu}. Here one may focus on the top two panels. The plot format and choice of symbol sizes and line widths make precise comparisons quite difficult. However, close attention to detail reveals that VISHNU predicted proton and Lambda trends fall in the wrong order on \pt, with Lambdas to the left of protons (intercepts at lower \pt), thus not following ``mass ordering'' which should be a basic outcome of hydro theory. The paper conclusion does admit that VISHNU fails to describe data, especially for more-central collisions where a hydrodynamic description should be most appropriate according to a conventional flow-QGP narrative.

\section{Conclusions} \label{conclude}

Taken together the results of this study, combined with related previous findings, appear to indicate that the azimuth quadrupole particle source is a {\em distinct particle production mechanism} with unique characteristics that may be derived from $v_2(p_t,n_{ch})$ data in combination with other results. The nature of the source mechanism is suggested by results for 200 GeV \pp\ collisions: a QCD three-gluon interaction. A quite similar data trend for \pbpb\ collisions suggests that the mechanism is universal within high-energy A-B collisions. Given overall data trends it is unlikely that the quadrupole component includes most hadrons emerging from Hubble-like expansion of a bulk medium as is conventionally assumed. It is more likely that the quadrupole component is ``carried'' by a small minority of final-state hadrons. Hydrodynamic theory descriptions appear to be falsified by $v_2(p_t,n_{ch},\sqrt{s_{NN}})$ data as demonstrated herein.




\end{document}